\documentclass{article}
\usepackage{wds11,epsf}
\usepackage{a4wide}
\usepackage[square,numbers]{natbib}
\usepackage[ruled,vlined]{algorithm2e}
\usepackage{graphicx}
\usepackage{color}
\usepackage{amsfonts}
\usepackage{multirow} 
\usepackage{booktabs}

\newcommand{\beq}{\begin{equation}}
\newcommand{\eeq}{\end{equation}}

\lefthead{Ad\'{a}mek et al.}
\righthead{The Implementation of a Real-Time Polyphase Filter}

\begin{document}

\title{The Implementation of a Real-Time Polyphase Filter}

\author{Karel Ad\'{a}mek and  Jan Novotn\'{y}}

\affil{Institute of Physics, Faculty of Philosophy and Science, Silesian University in Opava, Bezru\v{c}ovo n\'{a}m. 13, CZ-74601 Opava, Czech Republic}

\author{ Wes Armour}
\affil{Oxford e-Research Centre, University of Oxford, 7 Keble Road, Oxford OX1 3QG, United Kingdom}


\begin{abstract}
In this article we study the suitability of different computational accelerators for the task of real-time data processing. The algorithm used for comparison is the polyphase filter, a standard tool in signal processing and a well established algorithm. We measure performance in FLOPs and execution time, which is a critical factor for real-time systems. For our real-time studies we have chosen a data rate of 6.5GB/s, which is the estimated data rate for a single channel on the SKAs Low Frequency Aperture Array. Our findings show that GPUs are the most likely candidate for real-time data processing. GPUs are better in both performance and power consumption.
\end{abstract}

\begin{article}

\section{Introduction}
The complexity of sensor technology and computational equipment increases at ever greater rates. With this increase in complexity comes an increase in data production. Various large scale scientific experiments, in many different fields, now produce vast amounts of data; due to the volume of such data it is typically impractical to store all of the data created by such experiments. 
For successful scientific outcomes from such experiments important events must be first detected and then captured, to do this without storing the data produced by the experiment the data must be processed in real-time. 
The processing of incoming data is seldom a single step process, typically involving many different operations that have to be performed in a particular order on the received data. These operations together form a data processing pipeline which performs not only signal analysis and filtering, but also has the important task of screening data for scientifically interesting events, these events, when detected are stored, and the rest of the data can be discarded.

Real-time data processing is a phenomenon which, we believe, will become more prominent in the near future as it allows us to interrogate data as it is captured and negates the need to store large amounts of data at high-speed.
Due to the computational demands of real-time data processing parallel hardware like many-core GPU cards or Xeon Phi accelerator cards are required to accomplish this task.

This article focuses on a single data processing step in such a pipeline which is used on cutting edge radio-telescope experiments. This step is called the polyphase filter (PPF).
The polyphase filter has two major functions. It serves as a bandpass filter and secondly it suppresses the drawbacks of a discrete Fourier transformation (DFT), namely DFT leakage and scalloping loss. Here we investigate the PPF on different computational platforms (GPU, CPU, Xeon Phi) and use different programming techniques to implement it. 

Each of our implementations will be available in the Astro-Accelerate library. The Astro-
Accelerate project focuses on the implementation of real-time processing modules for time-domain radio-astronomy experiments on computational accelerators such as GPUs or Xeon Phi. 

In this article we discuss the comparison of programming techniques for our platforms under investigation, we also compare achieved performance and achieved multiples of real-time for the PPF on our different hardware.

We have selected the predicted output of single SKA channel for SKAs Low Frequency Aperture Array as a model for our input data, which should give about 6.5GB/s of data.

The polyphase filter has previously been implemented for LOFAR \cite{Vel:2012:PFBLOFAR}, however this implementation is targeted specifically at LOFAR, while our implementation is more general. Moreover we add support for Xeon Phi, which has not been implemented.

\section{Polyphase filter}
The Polyphase filter (PPF) has two main purposes. Firstly the PPF serves as a bandpass filter, after application to the data the desired frequency range is kept, while rest is set to zero. Secondly the PPF suppresses the drawbacks of the application of a discrete Fourier transformation (DFT) to data. These drawbacks are DFT leakage and DFT scalloping loss.
In signal processing terms the PPF is a linear filter applied on $N$ frequency channels. These channels represent DFT frequency bins after DFT. 
There are two kinds of linear filters, which differ in their behavior (response) to a single pulse in the input data. 
The First group are called FIR (Finite Impulse Response) filters. As the name suggests the response of these filters to a single pulse in the input data is finite in time. The second group are called IIR filters, IIR is an abbreviation of Infinite Impulse Response. 

\begin{figure*}[ht]
  \begin{center}
	\resizebox{140mm}{!}{\includegraphics{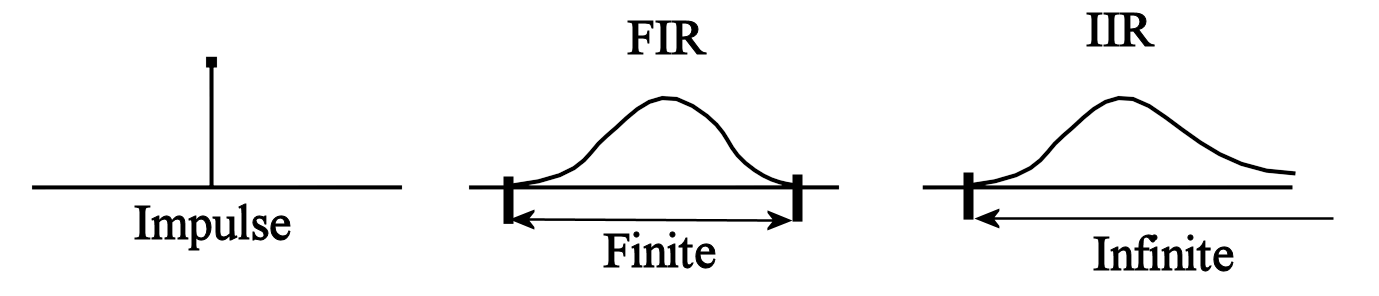}}
	\caption{\textit{This figure illustrates the behaviors of both groups of linear filters to a single pulse (left picture). The FIR filter's (middle) response is finite in time, while the IIR filter's (right) response is infinite in time.}}
    \label{FIR_IIR}
  \end{center}
\end{figure*}

These filters are recurrent and use their own output from previous calculations performed on a past samples to produce the current output which is formed from the latest sample. 
This property means that the response of these filters to a single pulse is infinite in time. An illustration of the respective behaviors of each of these types of filter is shown in the figure~\ref{FIR_IIR}. 

For our implementation of the PPF we have chosen the FIR filter, since they are stable against error in raw input data from the detector and they parallelize better. The stability of the FIR filter comes from their limited response in time. If we have some critical errors in our data, after using a FIR filter only part of our filtered data gets corrupted, this is opposed to IIR filter where all data after the error are affected. The FIR filter is also more easily parallelized since input to each filter is independent of any previous results of that filter and only raw input data are used to produce filtered data.

The FIR filter itself is a convolution of input time-domain data $x[n]$, usually in form of voltage data from the detector, and a response function $b_i$. The mathematical formula of such a FIR filter is given by \cite{LYONS:UDSP} as:
\begin{equation}
y[n]=\sum_{i=0}^{T-1}x[n-i]b_i,
\end{equation}
where $[\,]$ indicates that a physical quantity is discrete (sampled). The quantity $y[n]$ represents the filtered data, $x[n-i]$ means that the filter takes into account the most recent sample and $T-1$ past samples. The number of samples used is called taps. Quantities $y[n]$ and $x[n]$ are assumed to be complex.

The motivation behind the choice of coefficients $b_i$ is to enhance the frequency filtering nature of FIR filter. The choice of the coefficients $b_i$ depends on desired features of a FIR filter. We assumed a constant response function throughout the computation.

The ideal frequency filter is a rectangular shaped filter (or window) which would be multiplied by the input frequency data and would then output our desired frequency range.
However since we perform our filter on time domain data we require the inverse discrete Fourier transformation (IDFT) of our rectangular window, which is a $\textnormal{sinc}(x)=\sin(x)/x$ function. To calculate our coefficients we need the sampled version of the $\textnormal{sinc}$ function. The number of coefficients generated this way must be finite, thus we must cut off the tails of sampled $\textnormal{sinc}$ function at an appropriate point. This cut leads to an effect similar to DFT leakage.

At the moment we are using the Kaiser window (\cite{LYONS:UDSP}).

\section{Implementation}
We have targeted three different types of computational hardware to investigate; these are CPUs, Graphic cards (GPUs) and Xeon Phi. To compare the performance of each we have used an estimated data rate for a single channel from SKAs Low Frequency Aperture Array (LFAA), which is 6.5GB/s. Whilst it is likely that FPGAs (field-programmable gate array \footnote{More at http://www.altera.com/products/fpga.html}) will be the computational platform of choice for channelization of SKA data, computational accelerators such as GPUs and Xeon Phi have several advantages over FPGA devices. These are:
\begin{itemize}
\item{Faster to reconfigure -- to upload a new map to a FPGA takes of the order of seconds, whereas this takes microseconds for other accelerators.}
\item{Ease of programming/reprogramming: Traditionally accelerators are easily to program using C style languages or compiler declarations, giving them more flexibility (however we note that with the introduction of OpenCL for FPGAs this is now changing).}
\item{Software eco-system: Many accelerators have a large software eco-system to draw from.}
\end{itemize}

\subsection*{CPU}
Our CPU platform is a dual Intel Xeon E5-2650 configuration. The key in achieving good performance on a CPU is to utilize all processing cores present and to use the large vector processing units (VPUs) that modern CPUs utilize to gain high processing performance.
The vector units execute Single Instruction Multiple Data (SIMD) operations. The number of operations performed simultaneously by each VPU depends on the generation of the CPU.
Our CPUs have AVX units\footnote{More at https://software.intel.com/en-us/articles/introduction-to-intel-advanced-vector-extensions} which have 256 bit wide registers for SIMD operations. This means these registers hold 8 single precision floating point numbers simultaneously. In order to make use of the SIMD operations we used the Intel intrinsic instructions. These instructions provide a convenient way to access assembly instructions that can be executed on the VPU. For parallelization across cores we have used OpenMP. More about CPU optimizations and caches can be found in  \citep{Dre:2007:Wepsnam}. 
\subsection*{Xeon Phi}
The Xeon Phi is a new computational accelerator produced by Intel. The Phi is derived from CPU cores and thus shares lot of similarities with the CPU. The Xeon Phi has wider SIMD registers (512 bit) and many more cores. In this work we have used Xeon Phi 5110P, this has 60 cores, each core can run up to four threads giving in total 240 threads. Porting existing CPU code to Xeon Phi is relatively straight forward due to the similar programming model. General code can be run directly in native mode, assuming input and output are changed to suite Xeon Phi. Porting code with intrinsic instructions is more problematic since by using intrinsic instructions one binds the code to a specific CPU generation but also imposes constrains on data division. This was proven to be valid by \citep{Sch-Ule:2012:EarlyExp} on wide range of codes or by \citep{Cra-Sch:2012:Openmp} when they investigated performance of Sparse-Matrix-Vector-Multiplication. 
\subsection*{GPU}
For our GPU investigations we have chosen graphic cards produced by NVIDIA and have looked at their last three generations of GPU. From the Fermi generation we have investigated the GTX 580 gaming card, from the Kepler generation a scientific card, the K40, and from the Maxwell generation we have used a power efficient GPU, the GTX 750Ti. To obtain performance on GPUs one must still divide data as one would for CPU SIMD operations, but one must also be mindful of the specifics of the GPU platform. More about GPU programming can be found at \citep{NVIDIA:Prog}. 

Technical parameters of all used platforms can be seen in table 1\ref{hardware}. 

\begin{table}
	\centering
	\caption{Specification comparison of our investigated many-core platforms. }
	\begin{tabular}{lrrrrr}
	\toprule
	\multirow{2}{*}[-2pt]{Platform} & Frequency & Memory bandwidth & Peak Performance & TDP & Price\\
	& (GHz)& (GB/s)    & (GFLOP/s)  & (W) & (\$) \\
	\midrule
	Xeon E5-2650 2x   & 2.00 & 102 & 512  & 380 & 4448\\
	Xeon Phi 5110P    & 1.05 & 320  & 1920 & 225 & 2650\\
	Fermi GTX 580     & 1.63 & 196  & 1669 & 244 & 499\\
	Kepler Tesla K40  & 0.88 & 288  & 5045 & 235 & 5499\\
	Maxwell GTX 750 Ti & 1.25 & 86.4 & 1605 &  60 & 149\\
	\bottomrule
	\end{tabular}
	\label{hardware}
\end{table}

\begin{algorithm}
 \SetAlgoLined
	\SetKwData{nChannels}{nChannels}\SetKwData{nTaps}{nTaps}\SetKwData{nSpectra}{nSpectra}
	\SetKwData{Coeff}{coeff}
	\SetKwData{Data}{input\_data} \SetKwData{Spectra}{spectra}
	
	\SetKwFunction{MLoad}{\_mm\_loadu\_ps}\SetKwFunction{MSetZ}{\_mm\_setzero\_ps}
	\SetKwFunction{MMul}{\_mm\_mul\_ps}	\SetKwFunction{MAdd}{\_mm\_add\_ps}
	\SetKwFunction{MSto}{\_mm\_store\_ps}
	\SetKwInOut{Input}{input}\SetKwInOut{Output}{output}

	\For{$bl=0$ \KwTo \nSpectra}{
		\For{$c=0$ \KwTo \nChannels}{
			\For{$t=0$ \KwTo \nTaps}{
				\Spectra = \Coeff * \Data + \Spectra\;
			}
		}
	}
 
 \caption{Serial implementation of PPF filter.}
\label{alg:simplePFB}
\end{algorithm}

To perform the DFT we have used third party FFT libraries. For the CPU and the Xeon Phi we used the Intel MKL library and for GPUs we used the cuFFT library provided by NVIDIA.

\begin{figure}[hdt]
  \begin{center}
      \includegraphics{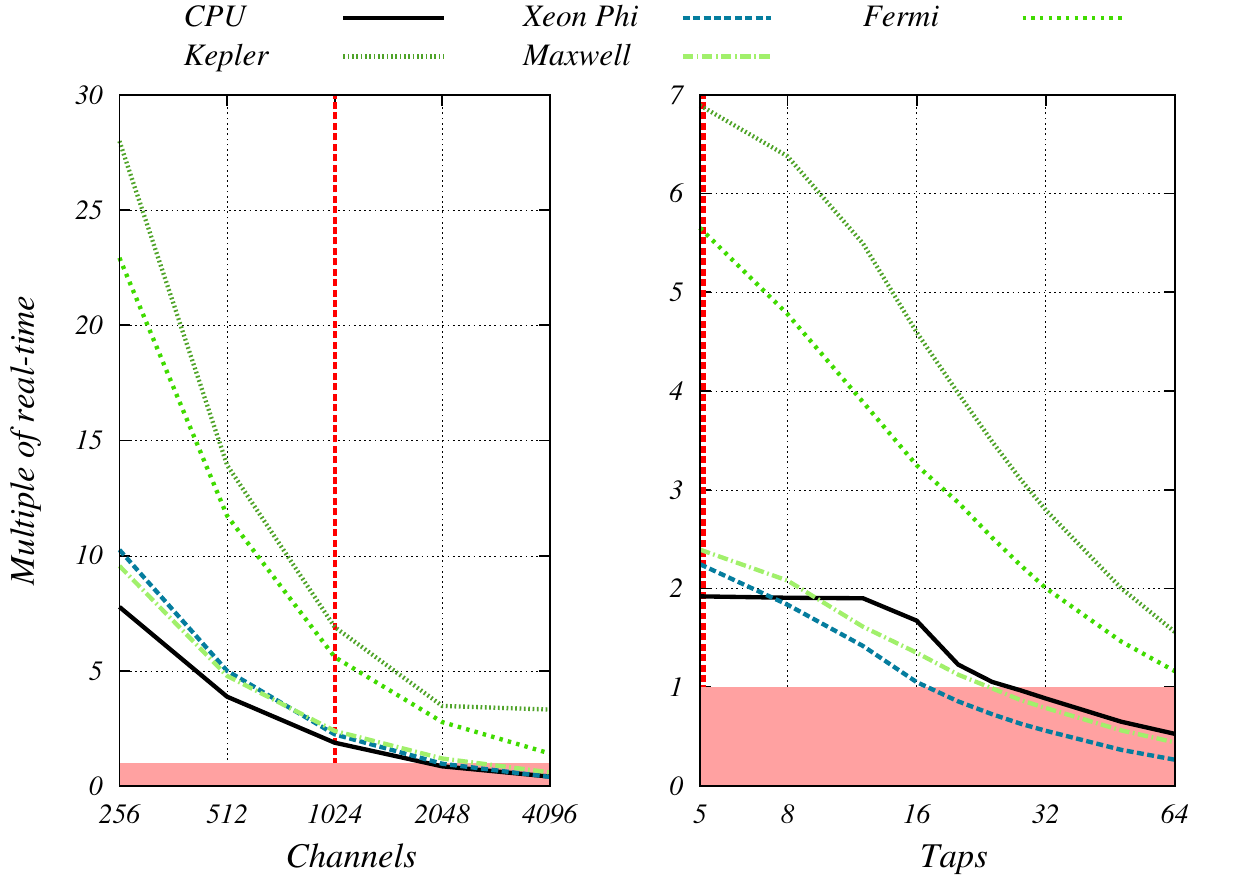}
    \caption{\textit{Multiples of real-time $M_c$ for various platforms and data configurations.}}
    \label{fig:mcresults}
  \end{center}
\end{figure}

We have chosen to compare our computational platforms based on a measure of multiples of real-time. By this we mean that we consider the time taken to process the amount of data contained in a single second of our data stream, for our example data this would be 6.5GB. Hence our multiples of real-time metric is defined as:
\beq
N=\frac{\textnormal{one second}}{\textnormal{time taken to process}}.
\eeq
So for example if the computational platform can compute one second of data in 0.5 of a second then the multiple of real-time value will be $M=2$. 
We must also separate this quantity into two different measures of multiples of real-time, we call these values $M_b$ and $M_c$. The first $M_b$ measures the ability of the platform to compute our chosen amount of data with all of the associated data transfer times included. This is very important for computational accelerators because we have to transfer data via the PCIe bus, which connects them to the host computer. The quantity $M_c$ discards the PCIe transfer time and focuses on computational potential of the accelerator. The reason we include this value is that within a streaming pipeline it is highly probable that the data would already be present in accelerator memory due to previous operations and would continue to reside in accelerator memory for further operations. Hence the PCIe transfer time in a streaming pipeline can only be sensibly measured as a fraction of the timing for the whole pipeline. In the case of our measurement $M_c$ we assume that data is already present in the accelerator device memory (GPU, Xeon Phi). For the CPU case it holds that $M_b=M_c$. We present measured values for $M_b$ in Table 2 and measured values for $M_c$ in the figure \ref{fig:mcresults}.

The results from Table 2 need more explanation. Although all devices are capable of real-time processing in the example we present, if we consider increasing the number of channels or the time sampling this would no longer be true. Along with this the PFB is often used in conjunction with other processing modules so has to run greater than real-time to accommodate them. When looking at our results from Table 2 it is also important to consider data movement in a real-time streaming pipeline. Whilst we consider movement over the PCIe bus, we do not consider the data movement from the detector or upstream digital signal processor to the host device. In this scenario the data will likely travel over the PCIe bus to be buffered in RAM. This isn't considered in our CPU results.

  \begin{center}
\begin{table}
	\caption{Comparison of performance in two execution modes. In the single mode regime the computational accelerator computes only the polyphase filter, thus transfer times needed for sending data to and from the accelerator must be included. This regime is represented by $M_b$ and by the mean utilization of PCIe Bus. In streaming mode we assume that data are already present on the device, thus the transfer times are omitted. For both modes we used our SKA estimate of 6.5GB/s.}
	\begin{tabular}{lrrrr}
	\toprule
	\multirow{2}{*}[-2pt]{Platform} & \multicolumn{3}{c}{Single mode} \\
	\cmidrule(r){2-4} & $M_b$ (1/s) & \multicolumn{2}{c}{PCIe usage (GB/s)  (GB/s)} \\
	\midrule
	Xeon E5-2650       & 1.92 & \multicolumn{2}{c}{---}       \\
	Xeon Phi 5110P     & 0.40 & \multicolumn{2}{c}{5.2 (33\%)}   \\
	Fermi GTX 580      & 0.37 & \multicolumn{2}{c}{4.8 (30\%) }  \\
	Kepler Tesla K40   & 1.38 & \multicolumn{2}{c}{17.9 (57\%)}  \\
	Maxwell GTX 750 Ti & 0.82 & \multicolumn{2}{c}{10.7 (34\%)}  \\
	\bottomrule
	\\
	\toprule
	\multirow{2}{*}[-2pt]{Platform}  & \multicolumn{3}{c}{Streaming mode}\\
	\cmidrule(r){2-4} & $M_c$ (1/s) & GFlops (GFlops/s) & Bandwidth (GB/s) \\
	\midrule
	Xeon E5-2650       & 1.92 & 233 (23\%) & 50  (49\%) \\
	Xeon Phi 5110P      & 2.25 & 269 (14\%) & 59  (18\%) \\
	Fermi GTX 580     & 5.65 & 657 (39\%) & 148 (76\%) \\
	Kepler Tesla K40  & 6.89 & 801 (16\%) & 181 (63\%) \\
	Maxwell GTX 750 Ti  & 2.39 & 278 (17\%) & 63  (73\%) \\
	\bottomrule
	\end{tabular}
	\label{tab:results}

\end{table}
  \end{center}

\section{Conclusions}
We have produced many-core accelerated implementations of the PPF for three generations of NVIDIA GPU and for Intel Xeon Phi. We have also produced a multi-core version for CPUs that makes use of the large VPUs present on modern CPUs. We have compared results from these codes on the stated platforms.
Our results show that from the perspective of performance and power efficiency the GPU prevails, however the limiting factor for the GPU is the PCIe bus. From a project perspective Xeon Phi prevails in terms of overall cost because one does not need to learn the intricacies of GPU programming, the Phi can be accessed quite easily by using the tools and skills one knows from the established CPU programming paradigm or by simply adding compiler directives. The transition from CPU code to Xeon Phi code is a relatively easy transition to make; this is advantageous if there is existing CPU code, making the porting of the code very easy. Our CPU results are also worthy of consideration. Our best implementation for the PPF on the CPU code was only 3.6 times slower than the best performing Kepler K40 result. ($M_c(GPU)=6.89\, \textnormal{s}^{-1}$ versus $M_c(CPU)=1.92\, \textnormal{s}^{-1}$)

\section{Acknowledgements}
This work was carried out under projects Astro-Accelerate and ARTEMIS, which performs surveys for astrophysical transients using the LOFAR radio telescope. We would like to express our gratitude to the internal grants of the Silesian University in Opava FPF SGS/23/2013 and SGS/11/2013 and the Institutional support of the Silesian University in Opava. 

The authors further acknowledge the project Supporting Integration with the International Theoretical and Observational Research Network in Relativistic Astrophysics of Compact Objects, reg. no. CZ.1.07/2.3.00/20.0071, supported by Operational Programme \textit{Education for Competitiveness} funded by Structural Funds of the European Union and state budget of the Czech Republic. 

The authors would like to acknowledge the use of the Advanced Research Computing (ARC) in carrying out this work and OeRC developers computer and Oxford CCoE (Cuda Centre of Excellence) for providing nVidia cards. 

We would like to express thanks to the following people for support and helpful discussions: Zden\v ek Stuchl\'ik, John Miller, Stanislav Hled{\'i}k  and Aris Karastergiou.


\end{article}
\end{document}